\documentclass[onecolumn,superscriptaddress]{revtex4}

\usepackage{amsfonts}
\usepackage{amsmath}
\usepackage{amssymb}
\usepackage{graphicx}
\usepackage{dcolumn}
\usepackage{dsfont}
\usepackage{times}
\usepackage{epsfig}
\usepackage[]{units}

\usepackage[colorlinks=true,citecolor=blue]{hyperref}

\def\T{\tau} 
\def\Y{Z}
\def\cd{c} 
\def\sound{\sigma} 

\def\bse{\begin{subequations}}
\def\ese{\end{subequations}}

\def\dn{\mathop{\rm dn}\nolimits}
\def\sn{\mathop{\rm sn}\nolimits}

\newcommand{\R}{{\mathbb R}}

\def\sign{{\mathrm{sign}}}

\newcommand{\pa}{\partial}

\def\epsilon{\varepsilon}

\def\beq{\begin{equation}}
\def\eeq{\end{equation}}

\def\dn{\mathrm{dn}}

\begin{document}

\title{Integrable Approximations of Dispersive Shock Waves of the Granular Chain}

\author{Christopher Chong}
\affiliation{Department of Mathematics, Bowdoin College, Brunswick, Maine 04011}

\author{Ari Geisler}
\affiliation{Department of Applied Mathematics, University of Colorado, Boulder, 80309}

\author{Panayotis G. Kevrekidis}
\affiliation{Department of Mathematics and Statistics, University of Massachusetts, Amherst, Massachusetts 01003}

\author{Gino Biondini}
\affiliation{Department of Mathematics, State University of New York at Buffalo, Buffalo, New York 14260}

\begin{abstract}
In the present work we revisit the shock wave dynamics in a granular
chain with precompression. By approximating the model by an
$\alpha$-Fermi-Pasta-Ulam-Tsingou chain, we leverage the connection of the
latter in the strain variable formulation to two separate integrable models,
one continuum, namely the KdV equation, and one discrete, namely the Toda
lattice. We bring to bear the Whitham modulation theory analysis of such
integrable systems and the analytical approximation of their dispersive shock
waves in order to provide, through the lens of the reductive connection to
the granular crystal, an approximation to the shock wave of the 
granular problem. A detailed numerical comparison of the original granular
chain and its approximate integrable-system based dispersive shocks
proves very favorable in a wide parametric range. The gradual deviations
between (approximate) theory and numerical 
computation, as amplitude parameters of the solution
increase are quantified and discussed.
\end{abstract}

\date{\today}

\vspace{1cm}

\maketitle

\section{Introduction}

Granular chains consist of closely 
packed arrays of particles that interact elastically upon compression. They
have received much recent attention due to their potential in applications
(such as shock absorption, frequency conversion, and energy harvesting),
recent advances in experimental platforms (including heterogeneous and random ones, ones involving mass-in-mass, mass-with-mass, branching, and intruder-based ones)
and the mathematical richness of the underlying equations. See
~\cite{Nester2001,granularBook,yuli_book,gc_review,sen08} for comprehensive reviews
on granular chains. From a fundamental perspective, there are three structures
of granular chains of central importance: The solitary wave, the breather, and the dispersive shock
wave. While the solitary wave and breather have been studied extensively in 
the context of such nonlinear lattices, its dispersive shock
waves (DSWs) are far less understood.

A DSW connects states of different amplitude 
via an expanding modulated wave train. The study of DSWs in spatially continuous media has been an active area of research since Whitham's seminal work \cite{Whitham74} over 50 years ago.
There has been, however, a renewed excitement concerning DSWs. This has largely been inspired by groundbreaking experiments observing DSWs in ultracold gases, optics, superfluids, electron beams, and plasmas \cite{dsw,Hoefer2016,Trillo2018}. The new body of mathematical work is summarized in recent review articles on the subject \cite{scholar,Mark2016,dsw}. Dispersive shock waves in one-dimensional (1D) nonlinear lattices (to be called lattice DSWs) have been explored numerically, and even experimentally in several works \cite{first_DSW,Nester2001,Hascoet2000,Herbold07,Molinari2009,shock_trans_granular,HEC_DSW}. 
Although much of the above motivation stems from the granular chains, it is of broad physical interest, as similar
structures have been experimentally observed, e.g., in nonlinear
optics of waveguide arrays~\cite{fleischer2}.
It is important to also highlight in this context 
another setup that has recently emerged, namely tunable magnetic
lattices~\cite{talcohen}. Here, ultraslow shock waves can arise and have 
been experimentally imaged.

The primary tool to analytically describe DSWs is the so-called Whitham modulation theory \cite{Whitham74,GP73,Karpman}.
In this framework, one derives equations describing slow modulations of the underlying parameters of a periodic wave by, for example, averaging 
the Lagragian action integral over a family of periodic wave trains \cite{dsw,Mark2016}. 
The existence of periodic waves has been proved \cite{Iooss2000,Pankov05,Herrmann10b} and corresponding 
 modulation equations have been derived \cite{Venakides99,DHM06}.
Explicit forms of the periodic waves may, however, not be 
available, resulting in modulation equations that are difficult to work with. 
Thus, in order to get a better understanding of DSWs beyond numerical
computations, it is useful to explore additional avenues beyond direct use of the modulation equations. For instance, in \cite{marchant2012} analytical techniques to estimate the leading and trailing amplitudes are described. Recently in \cite{CHONG2022133533} a discrete conservation law was studied via a reduction to planar ODE dynamics. 

In this work, we will revisit the central and experimentally tractable
problem of DSW formation in the prototypical setting of granular 
crystals~\cite{first_DSW,Nester2001,Hascoet2000,Herbold07,Molinari2009,shock_trans_granular,HEC_DSW}. In the past, and for different structures of this
system, such as the traveling waves, resorting to continuum (such as
the Korteweg-de Vries (KdV) equation) or discrete (such as the Toda lattice)
integrable and hence analytically tractable limits has proven to be of
considerable value~\cite{Shen}. It is indeed that route that we will
examine herein for the realm of dispersive shock waves.
We will explore, in particular, how to exploit existing knowledge on DSWs in those systems to better understand DSWs of the granular chain, but also
to explore the extent of the validity of those approximations. Their validity should not be taken
for granted, since, e.g., in the case of the KdV equation, the long wavelength assumption needed for its derivation
is violated in the case of step initial data in which DSWs arise.

Our presentation will be structured as follows. In section II we will briefly
describe the underlying discrete granular problem. Then, in section III, we will
delve into the DSW description for the KdV model and its connection/comparison
with the granular one. In section IV, we will perform the corresponding
analysis and comparison in the case of the Toda lattice. Finally, in section V,
we will summarize our findings and present our conclusions and some possible
directions for future studies.

\section{Model Equations for the Granular Chain}

An idealized model of the monomer granular chain is given by~\cite{Nester2001,granularBook,yuli_book,gc_review,sen08}
\begin{equation}
	M \ddot{u}_n = \gamma [d_{0}+u_{n-1}-u_n]_+^{p}-
	\gamma [d_{0}+u_n-u_{n+1}]_+^{p},
\label{eq:model}
\end{equation}
where $M$ is the effective mass of each node and  $u_n=u_n(t)\in\R$ represents the displacement
of node $n$ from its equilibrium position at time $t\in\R$.  In the case of spherical particles
the nonlinear exponent is $p=3/2$ (other exponents are also possible depending on the geometry,
contact angle, and even material type \cite{nesterenko07}).
The parameter $d_0$ represents
a static displacement (the so-called precompression) that allows additional tunability in the degree of nonlinearity (where $d_0 = 0$ represents a purely nonlinear force
and $|d_0/x| \gg 1$ represents a nearly linear force).
The brackets account for the fact that there is no force in the absence
of contact, i.e. $\lbrack x \rbrack_+ = \max(0,x)$.  
While Eq.~\eqref{eq:model} ignores effects such as
dissipation, it has proven to be a reliable model when compared against experiments \cite{Nester2001,HEC_DSW,origami2,moleron}.

In the case of nonzero precompression ($d_0\neq0$), a Taylor approximation of the inter-particle
force $\gamma(d_0-x)^p$ can be used, which, when neglecting all terms beyond cubic powers, reduces the granular chain model to
the Fermi-Pasta-Ulam-Tsingou model \cite{FPU55} of the form:
\begin{equation} \label{eq:fput_granular}
	\ddot{u}_n = K_2 (u_{n-1} - 2 u_n + u_{n+1}) + 
                 K_3\left[ (u_{n+1} - u_{n})^2 - (u_{n} - u_{n-1})^2   \right] + 
                       K_4\left[ (u_{n+1} - u_{n})^3 - (u_{n} - u_{n-1})^3   \right] 
\end{equation}
where
$$K_2 =\frac{p \gamma d_0^{p-1}}{M}, \,K_3 = -\frac{p(p-1)}{2M} \gamma 
d_0^{p-2},\, K_4 = \frac{p(p-1)(p-2)}{6M}\gamma d_0^{p-3}.$$
This approximation assumes that the strain $y_n = u_n-u_{n+1}$ is small relative to
the precompression, i.e. $|y_n| \ll d_0$ for all $n$. Thus, oscillations should not exceed the overlap caused by the precompression, suggesting the particles will remain in contact, hence the dropping of the bracket notation. For the graphs in this paper,
we will plot the strain $y_n$, since the size of $y_n$ will tell us how close 
we are to the assumption $|y_n| \ll d_0$. With this definition
of the strain, waves with $y_n >0$ imply the beads
are squeezed beyond the precompression amount, and are in this sense compression waves. On 
the other hand, waves with
$y_n < 0$ imply the beads  are squeezed less than the precompression amount. While the beads may still be physically compressed, waves with $y_n < 0$
can be thought of as a type of tensile wave since they are ``stretched" relative to the precompression amount.  If $ y_n < -d_0$ then the beads at the lattice
sites $n$ and $n+1$ lose contact, and the nonlinearity induced by the bracket
$\lbrack x \rbrack_+$ is required. The strongly nonlinear cases of $y_n \gg d_0$
and $y_n < -d_0$ will not be covered by the analysis in this paper, which deals
with the weakly nonlinear limit. Since the granular chain will necessarily have $K_3<0$, we can already expect the appearance of compression type waves. Other types of lattices,
such as strain-softening ones~\cite{RarefactionNester}, will have $K_3 >0$, and will thus 
generate tensile waves. The analysis in this paper 
could be applied to other
strain-hardening lattices (where the relevant Taylor expansion has $K_3 < 0$),
such as magnetic lattices~\cite{magBreathers}, but not strain-softening ones
(where the relevant Taylor expansion has $K_3 > 0$).
For all simulations we use the parameter values 
$\gamma = 2^{3/2}/3$, $M=1$, and $d_0 = 1/2$.
In this case, the Taylor coefficients
are $K_2 = 1$ and $K_3 = -1/2$. The reason for the choice
is convenience when comparing the granular lattice
to the Toda chain. Before discussing the Toda approximation,
we consider the KdV description, which is simpler.

\section{KdV Description of the Granular Chain}

The connection between the FPUT lattice and the KdV equation dates back to
the seminal work of Zabusky and Kruskal \cite{Zabusky}, and is one of the earliest in the study
of nonlinear waves. While many aspects of the FPUT lattice have been explored
through the KdV lens \cite{FPUreview,VAINCHTEIN2022} (see also~\cite{Shen}
in the context of traveling waves in granular crystals), 
its DSWs have gained less attention. One notable
work in this light is \cite{Paleari2006}, which explores the connection
of metastability and DSWs using the KdV equation.
The KdV equation
is derived using a small-amplitude and long wavelength assumption. While
the small-amplitude aspect will allow us to move from the granular chain model
Eq.~\eqref{eq:model} to the FPUT model of Eq.~\eqref{eq:fput_granular},
the long wavelength assumption is more troubling in the context of DSWs.
Riemann initial data clearly violate the long wavelength assumption, and so
it is not to clear to what extent, if at all, the KdV description of 
the granular DSWs will be valid. We explore exactly this question in this section.
Since we are reporting results in terms of the strain, it is natural to express
Eq.~\eqref{eq:fput_granular} in terms of the strain variable
$y_n = u_{n} - u_{n+1}$,
\begin{equation} \label{eq:fput_granular_strain}
\ddot{y}_n = K_2 (y_{n-1} - y_n + y_{n+1}) - 
                 K_3 ( y_{n-1}^2 -  2y_{n}^2 + y_{n+1}^2  )  
\end{equation}
We leave out the cubic (and other higher order) terms since they will not play a role in the analysis.
Upon substitution of the ansatz
\begin{equation}
	y_n = \epsilon^2 Y(X,T) , \qquad X=\epsilon(n-\sound t),\quad {T} = \epsilon^3 t 
\label{ansatz}
\end{equation}
into Eq.~\eqref{eq:fput_granular_strain} and Taylor expanding the $Y(X \pm \epsilon,T)$ terms,
one finds that the terms up to $\mathcal{O}(\epsilon^6)$ in the residual will be eliminated
if $Y$ satisfies the following KdV equation,
\begin{equation} \label{kdv}
\partial_T Y +   \frac{\sound }{24} \partial_X^3 Y - \frac{K_3}{\sound } Y \pa_X Y  = 0,
\end{equation}
where the sound speed $\sound $ is defined through $\sound^2 = K_2$. Note that,
since $K_3 < 0$, the solitary waves of the above KdV equation will 
have $Y>0$, i.e., the solitary waves of the granular chain are compression waves,
as expected. See \cite{Shen2014} for a detailed discussion of the KdV (and Toda)
description of the granular chain solitary waves.
Through the scaling $\T=\sound T /24$ and $\Y = -\frac{24 K_3}{\sound^2}Y$  
we can cast all KdV coefficients to unity, 
\begin{equation} \label{kdv_unity}
\partial_{\T}  \Y +  \partial_X^3 \Y + \Y \pa_X \Y = 0.
\end{equation}
Solutions of this KdV equation
that are of critical importance for our purposes are the periodic traveling  waves,
\begin{equation}\label{per}
\Y(X,\T) = r_1 + r_2 - r_3 + 2(r_3 - r_1) \dn^2 \left(\sqrt{\frac{r_3 - r_1}{6} }(X - V \T) ; m \right), \quad V = \frac{r_1 + r_2 + r_3}{3}, \quad m = \frac{r_2-r_1}{r_3-r_1}
\end{equation}
which are parameterized by $r_1,r_2,r_3$, and dn is one of the Jacobi elliptic functions with elliptic parameter $0 \leq m \leq 1$
\cite{NIST2010}. 
Notice that the limit of these waves
as $m \rightarrow 1$ are the solitary waves of the model.

\subsection{DSWs of the KdV equation}

The DSWs of the KdV equation \eqref{kdv_unity} are well studied, and represent
a textbook example of DSWs.
In the seminal paper \cite{GP73},
the following
initial conditions for Eq.~\eqref{kdv_unity} were considered in the KdV case:
\begin{equation}\label{step2}
\Y(X,0 ) = \left\{
\begin{split}
1, \quad & X < 0 \\
0, \quad & X > 0
\end{split}
\right.
\end{equation}
Assuming the parameters of the periodic wave (e.g., the $r_j$) as varying slowly with respect to $X,\tau$ and averaging three of the conserved quantities of the KdV equation over a period yields a set of three Whitham modulation equations \cite{Whitham74}. 
In the case of self-similar solutions, $r_j=r_j(X/\tau)$,
these equations have the form $(S_j - X/\tau)r'(X/\tau)=0$, 
where the characteristic speeds $S_j$ are nonlinear functions of $r_1,r_2$ and $r_3$.
Thus, in this self-similar framework, that $r_j$ is either constant, or $X/\tau$ is equivalent to the characteristic speed $S_j$.
Assuming step initial data of the form of Eq.~\eqref{step2}
one finds \cite{GP73},
$$r_1 = 0, \qquad r_2 = m, \qquad r_3 = 1. $$
Substituting these expressions into Eq.~\eqref{per}, one obtains the following formula for the
KdV DSW,
\begin{equation}\label{eq:kdv_dsw}
\Y(X,\T) = 2 \dn^2\left(\sqrt{\frac{1}{6}}\bigg(X - \frac{1+m}{3} \T\bigg) ; m\right) - (1 - m)
\end{equation}
where $X,\T$ are parameterized by $m$ through the expression 
\begin{equation}
X/\T = S(m)\,,
\end{equation}
with
$S(m)$ the second characteristic velocity of the Whitham equations 
(where the subscript 2 was ignored for simplicity), namely
\begin{equation} \label{eq:rvel2}
  S(m) = \frac{1+m}{3} - \frac{2}{3} \frac{m(1-m) K(m)}{E(m) - (1-m)K(m)},
\end{equation}
and $K(m)$ and $E(m)$ are complete elliptic integrals of the first and second kind, respectively.
Details for the derivation of this expression can be found 
in \cite{Gurevich1987,Kamchatnov,Mark2016}. 
In Eq.~\eqref{eq:kdv_dsw}, the limit $m\rightarrow 0$ corresponds to the trailing, harmonic wave edge, 
while the limit $m\rightarrow 1$ corresponds to the leading, solitary wave edge.
The trailing edge speed ($S_-$) and leading edge speed ($S_+$)  are obtained as limiting values in Eq.~\eqref{eq:rvel2}. In particular
\begin{equation}
 \lim_{m\rightarrow 0} S(m) =S_- = -1, \qquad \lim_{m\rightarrow 1} S(m) = S_+ = 2/3
\end{equation}
Translating back to the granular chain variables, $y_n(t)$,  we obtain the following approximation
for the core of the granular chain DSW:
\begin{equation}\label{eq:granular_kdv_dsw}
y_n(t) =  - \cd \frac{\sound^2}{ K_3 }\left(   2\dn^2\left(\sqrt{4 \cd} \left( n-  \left(1 +   \frac{m+1}{3} \cd  \right)\sound t + \theta_0(n,t) \right); m\right) - (1 - m) \right)
\end{equation}
where $\cd = \epsilon^2 / 24$ is a small parameter, and the variables $n,t$ are parameterized 
by $m$ through the expression
\begin{equation}
  \frac{(n-\sound t)}{\sound \cd t} = S(m)\,, 
\end{equation}
with $S(m)$ still given by Eq.~\eqref{eq:rvel2}. 
Note Eq.~\eqref{eq:granular_kdv_dsw} contains a slowly modulated phase shift
$\theta_0(n,t)$ that is not accounted for by the leading order Whitham theory \cite{SIAP59p2162}.
We have included it here, as it will be treated as a fitting parameter to account for a phase
mismatch between theory and simulation. 

From the above expression we can write the trailing edge speed ($n/t=s_-^{\rm KdV}$) and leading edge speed $(n/t=s_+^{\rm KdV})$
in terms of the original granular chain variables,
\begin{subequations}
\label{e:spmkdv}
\begin{eqnarray}
s_-^{\rm KdV} &=& \sound +\sound\cd S_-  = \sound \left(1 - \cd\right) \\
s_+^{\rm KdV} &=& \sound + \sound\cd S_+  = \sound\left(1 + \frac{2}{3}\cd \right).
\end{eqnarray} 
\end{subequations}
Thus the leading edge speed is supersonic, as it necessarily 
features larger than the
sound speed $\sound$, whereas the trailing edge speed is subsonic. Since the parameter
$\epsilon$ is assumed to be small (and hence also so is $\cd$), the leading edge is traveling just
above the sound speed, and the trailing edge is slightly below. Using
the parameter values $K_2=1$ and $K_3 = -\frac{1}{2}$,
we have the following estimates for the trailing edge speed,
and mean $\bar{y}_-^{\rm KdV}$ and the leading edge speed
and amplitude $(a_+^{\rm KdV} )$,
\begin{equation}\label{eq:kdv_speeds_param}
s_-^{\rm KdV} = 1 - \cd, \qquad  \bar{y}_-^{\rm KdV} = 2c, \qquad s_+^{\rm KdV} = 1 + \frac{2}{3}\cd, \qquad a_+^{\rm KdV} = 4c.
\end{equation} 
We write the leading and trailing edge characteristics explicitly with this choice of parameters due to 
their intimate connection with the Toda predictions. This is also why in Eq.~\eqref{e:spmkdv} and~\eqref{eq:kdv_speeds_param}  we used the 
KdV superscript to distinguish between these approximations and the ones using the Toda lattice,
described in section \ref{sec:toda}.
\begin{figure}[t!]
\centerline{
   \begin{tabular}{@{}p{0.33\linewidth}@{}p{0.33\linewidth}@{}p{0.33\linewidth}@{} }
     \rlap{\hspace*{5pt}\raisebox{\dimexpr\ht1-.1\baselineskip}{\bf (a)}}
 \includegraphics[height=4.4cm]{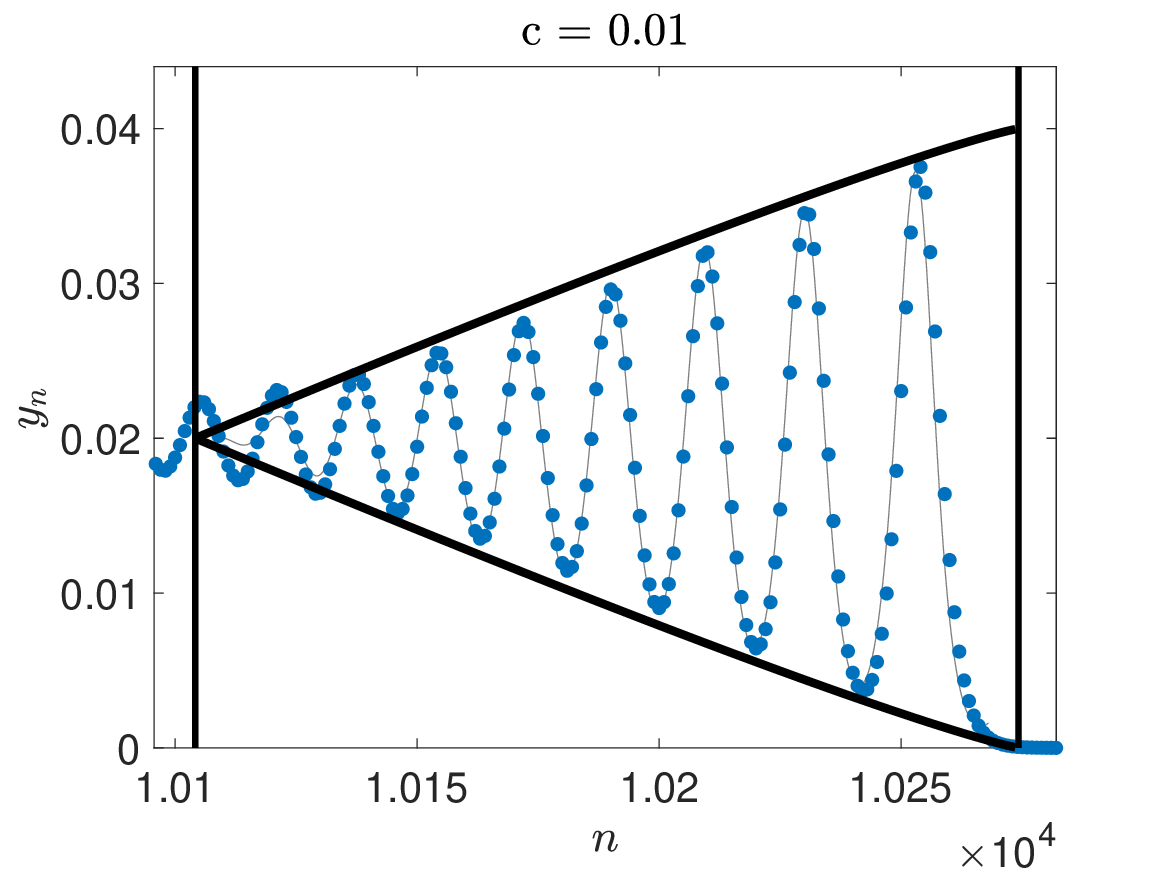} &
  \rlap{\hspace*{5pt}\raisebox{\dimexpr\ht1-.1\baselineskip}{\bf (b)}}
 \includegraphics[height=4.4cm]{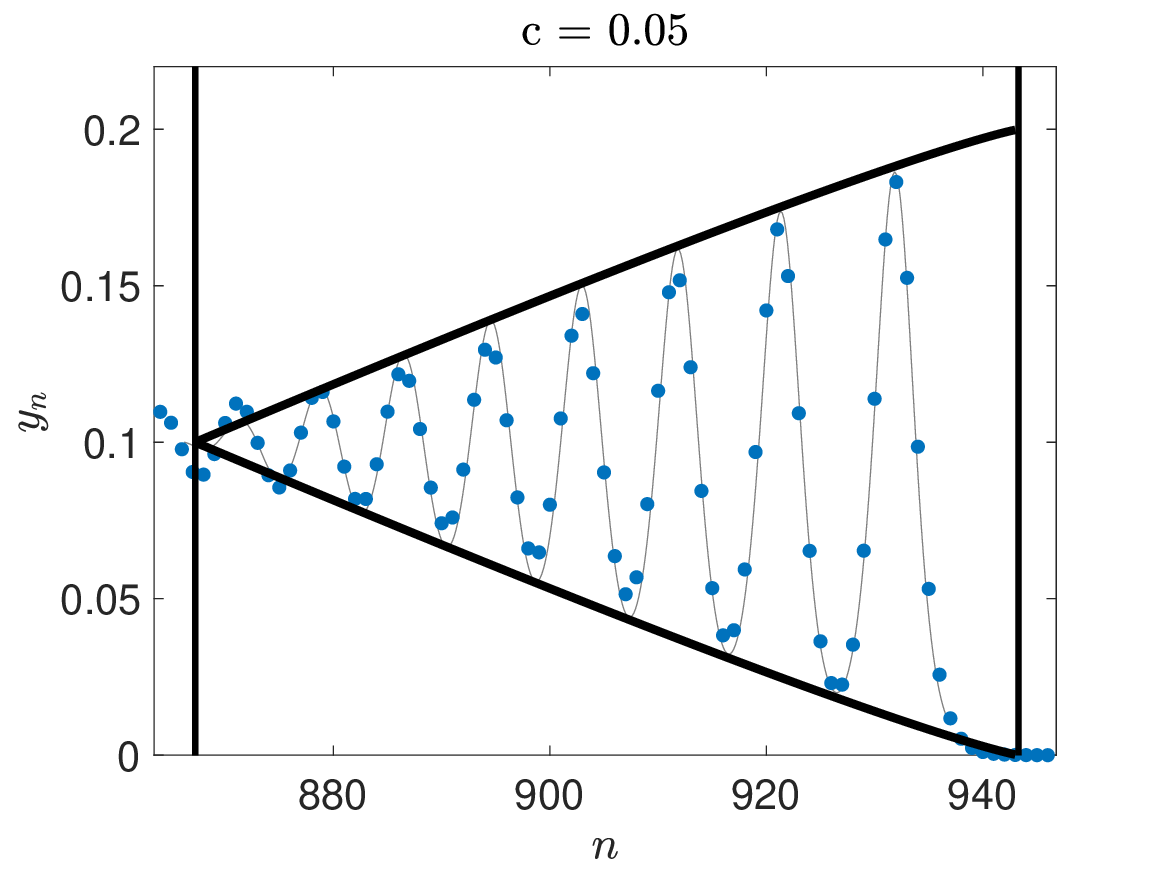} &
   \rlap{\hspace*{5pt}\raisebox{\dimexpr\ht1-.1\baselineskip}{\bf (c)}}
 \includegraphics[height=4.4cm]{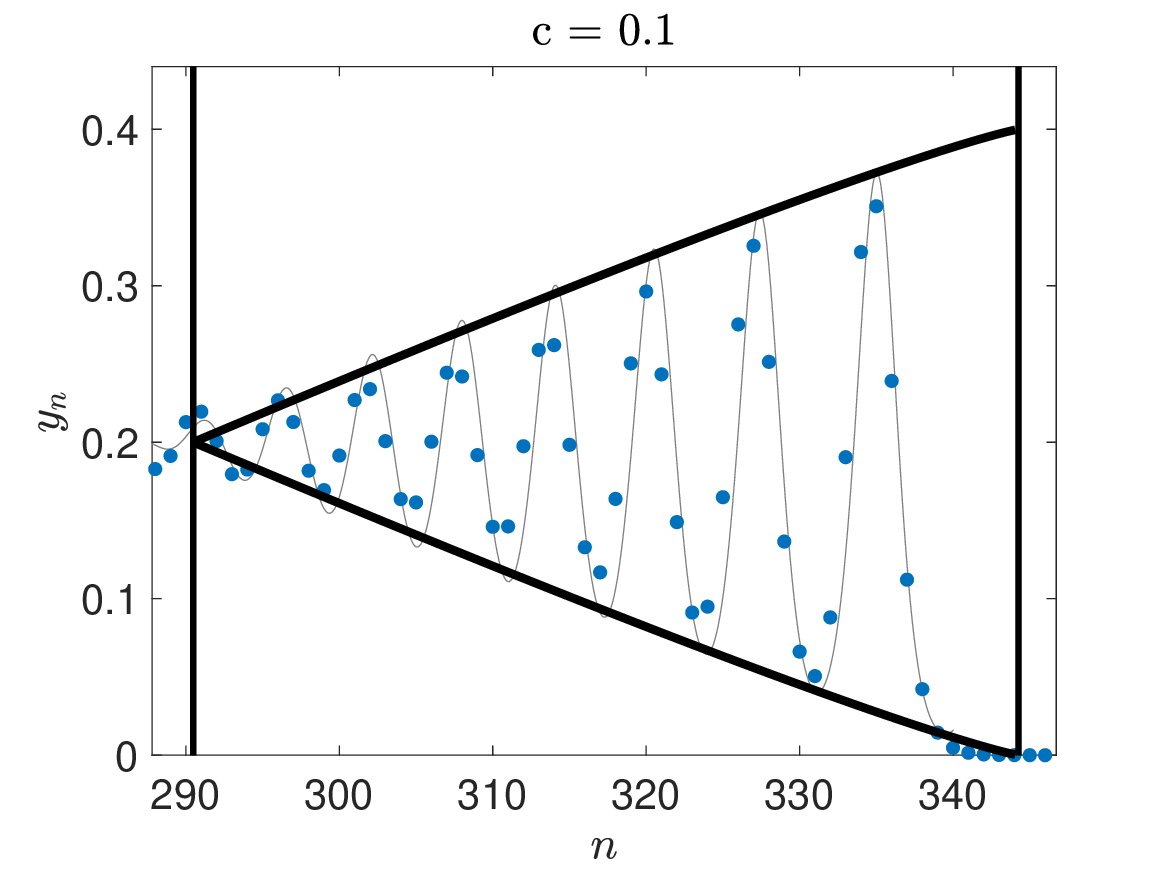} 
  \end{tabular}
  }
    \caption{Comparison of the KdV prediction and granular chain simulations in the strain variable, $y_n$, for various values of $\cd$. The KdV predictions, given by Eq.~\eqref{eq:granular_kdv_dsw}, are shown as solid lines, while
    granular chain simulations with initial data given by Eq.~\eqref{eq:initial_conditions}, are shown as markers.
    In all panels, the variables in the KdV scaling
    are fixed to $T=1200$ and $X\in[-50,30]$. The vertical lines
    are the predictions of the trailing edge ($n = s_-^{\rm KdV} t$) and leading edge ($n= s_+ t$).
    The sloped lines are the prediction of the envelopes.
    \textbf{(a)} The small parameter is $\cd = 0.01$ ($\epsilon \approx 0.49$) and the time in the original granular
    chain scaling is $t \approx 10,206$. The best-fit phase shift is the linear interpolation
    between $\theta_\ell = 5$ and $\theta_r = 7$.
    \textbf{(b)} $\cd = 0.05$ ($\epsilon \approx 1.1$), $t \approx 912$,
    $\theta_\ell = 0$ and $\theta_r = 1$.
     \textbf{(c)} $\cd = 0.1$ ($\epsilon \approx 1.55$), $t \approx 322$, $\theta_\ell = -2$ and $\theta_r = 0.5$.
    } 
    \label{fig:kdv_XTfixed}
\end{figure}

\subsection{Comparison of KdV and granular DSWs}

To write down the initial values for the simulation of Eq.~\eqref{eq:model} that correspond to Eq.~\eqref{ansatz}, we will need the
velocity
\begin{eqnarray*}
  \dot{y}_n(t) &=&  \frac{d}{dt} \epsilon^2 Y(X,T) \\
 &=& - \epsilon^3 \sound \pa_X Y + \epsilon^5 \left(  \frac{K_3}{\sound} Y \pa_X Y-\frac{\sound}{24}\pa_X^3 Y \right) 
\end{eqnarray*}
where we used the fact that $Y$ is assumed to solve
the KdV equation, Eq.~\eqref{kdv}. If we substitute $Y(X,0) = -\frac{\sound^2}{24 K_3} Z(X,0)$ into the above with $Z(X,0)$ defined in Eq.~\eqref{step2}, the spatial derivatives will be undefined at $X=0$. Thus, we replace
Eq.~\eqref{step2} with a smooth approximation of the unit step, namely,
\begin{equation}\label{step2smooth}
\Y(X,0 ) = \frac{1-\tanh(w X)}{2} =: f(X)
\end{equation}
in which case all quantities in the initial value are well-defined. Thus, the initial strain and velocity become,
\begin{subequations} \label{eq:initial_conditions} 
    \begin{gather} 
      y_n(0) =  -\frac{\epsilon^2 c_2^2}{24 K_3} f(\epsilon n) \label{eq:initial_strain} \\
      \dot{y}_n(0) = \frac{\epsilon^3 \sound^3}{24 K_3} f'(\epsilon n ) + \frac{\epsilon^5 \sound^3}{24^2 K_3}\left(  f(\epsilon n) f'(\epsilon n)+f'''(\epsilon n) \right)  \label{eq:initial_velocity} 
\end{gather}
\end{subequations}
For our first set of simulations, we initialize Eq.~\eqref{eq:model} with
Eqs.~\eqref{eq:initial_conditions} with $w=1/2$ and
for various values of the parameter $\cd$, see Fig.~\ref{fig:kdv_XTfixed}.
In this figure, we select fixed values of the KdV variables
$X \in [-50,30]$ and $T_f=1200 \, (\tau=50)$. For these values of the macroscopic parameters
the KdV DSW (see Eq.~\eqref{eq:kdv_dsw}) is developed, and features about 8 oscillations
within its core. To see the DSW in the granular chain develop to the same extent, we must simulate
until $t_f= \epsilon^{-3} T_f$. Since the leading edge is traveling at the speed $s_+$
the lattice must extend at least until $n= s_+^{\rm KdV} t_f$. For small values of $\epsilon$ (and hence $\cd$)
this leads to long simulation times with lattices that are quite large, see Fig.~\ref{fig:kdv_XTfixed}(a).
In this panel, we have $\cd=0.01$ ($\epsilon \approx 0.49$), which implies that $t_f = 10,206$
and that the largest lattice index should exceed $n=10,275$. The maximum strain is approximately $4c = 0.04$,
which is 8\% of the precompression amount $d_0=0.5$. This is a fairly weak nonlinear response, and one
would hope the KdV approximation to be accurate.  Indeed, by comparing the
solid lines (KdV approximation) and markers (granular chain simulation) of Fig.~\ref{fig:kdv_XTfixed}(a),
one sees good agreement. The sloped lines are the prediction of the envelope
of the DSW, which is also quite accurate. The vertical lines are the predictions of the trailing edge ($n = s_-^{\rm KdV} t$) and leading edge ($n= s_+^{\rm KdV} t$). In the figure, one can see linear
waves at the trailing edge that do no vanish, while the KdV prediction has vanishing
oscillations at the trailing edge. The existence of linear waves
at the trailing edge is well known in FPUT lattices.
While these linear waves will always be present, their amplitude
decays like $\sim t^{-1/3}$ \cite{MielkePatz2017} (and they are
not expected to be captured through Whitham theory). The
leading edge of the actual DSW is lagging behind the predicted location (given by $s_+^{\rm KdV} t$). 
This discprency becomes larger as $\cd$ is increased (compare panels (a)-(c)).
In general, there will be a phase mismatch between the theoretical prediction
and the actual granular DSW. To account for this, a phase shift $\theta_0(n)$ is applied
to the theoretical prediction by finding the best-fit phase shift at the trailing edge of the DSW
(to obtain a phase shift $\theta_\ell$) and at the leading edge (to obtain a phase shift $\theta_r$).
Then $\theta_0(n)$ is defined as the linear interpolation between these two shifts, namely 
$\theta_0(n) = \frac{\theta_\ell - \theta_r}{n_\ell - n_r}(n - n_r) + \theta_r$
where $n_\ell$ and $n_r$ are lattice indices close to the trailing and leading edge, respectively.
We practically used $n_\ell = s_-^{\rm KdV} t_f + 5$ and $n_r = s_+^{\rm KdV} t_f - 5$. Accounting for the phase
in this way results in good agreement between the KdV prediction and actual full profile of the DSW. Note, for reference purposes, that we do not shift the trailing and leading edge, nor the envelope predictions (thus the solid black lines are the ``original" prediction without
an empirically determined phase shift).
There is a number of causes for the phase mismatch between theory and actual DSW. The first
is that the initial condition does not lead to the immediate formation of a DSW, since
the smooth monotone decreasing initial data first needs to develop a gradient catastrophe.
Thus, one would expect the leading solitary wave to lag behind the prediction, due to
the later ``start" in the simulation. The catastrophe time in the derived KdV
equation can be approximated by computing when two arbitrary characteristics
lines of the underlying Hopf equation $\Y_\T + \Y \Y_X$ intersect, leading to the prediction
$\T_{\rm catastrophe} = -1/\min_\R f'(X)$ where $f(X)$ is  the initial datum
defined in Eq.~\eqref{step2smooth}. This could be used as a concrete correction for the phase shift. However, it is less obvious how to correct for the other two sources of the phase mismatch.
One is the missing phase description from the first order Whitham theory. While such a prediction
is possible in principle, the underlying complexity of the correction would undermine the elegant
simplicity of the approximation given in Eq.~\eqref{eq:granular_kdv_dsw}. Finally,
another source of mismatch will be in the inherent approximate nature of Eq.~\eqref{ansatz}.
While bounds for the error of the KdV approximation exist (assuming smoothness
of the underlying KdV solutions \cite{pegogf1,SW00}),
they cannot be used to correct the phase mismatch. Thus, we capture all the possible sources
of error in the phase by empirically determining the phase $\theta_0(n)$. Even without
this empirical correction, the leading and trailing edge, and the envelopes,
are well described by the KdV equation. Panels (b) and (c) of Fig.~\ref{fig:kdv_XTfixed}
are similar to panel (a), but consider larger values of the parameter $\cd$,
in order to test the practical limits of the approximation. In panel (b) we have
$\cd = 0.05$ ($\epsilon \approx 1.1$), which leads to a maximum strain of about
$4c = 0.2$, which is 40\% of the initial precompression. This is a fairly nonlinear
response, and indeed, the parameter $\epsilon \approx 1.1$ actually exceeds
unity, which is in clear violation of the smallness assumption. Nonetheless,
the KdV approximation is still quite reasonable. In panel (c) we have
$\cd = 0.10$ ($\epsilon \approx 1.55$), which leads to a maximum strain of about
$4c = 0.4$, which is 80\% of the initial precompression. For such large values
of the strain, one can start to see discrepancies between the KdV prediction
and granular chain dynamics, even after accounting for the phase mismatch.
From a qualitative perspective, however, the agreement remains reasonable
considering how large the strain is. 
\begin{figure}
\centerline{
   \begin{tabular}{@{}p{0.33\linewidth}@{}p{0.33\linewidth}@{}p{0.33\linewidth}@{} }
     \rlap{\hspace*{5pt}\raisebox{\dimexpr\ht1-.1\baselineskip}{\bf (a)}}
 \includegraphics[height=4.4cm]{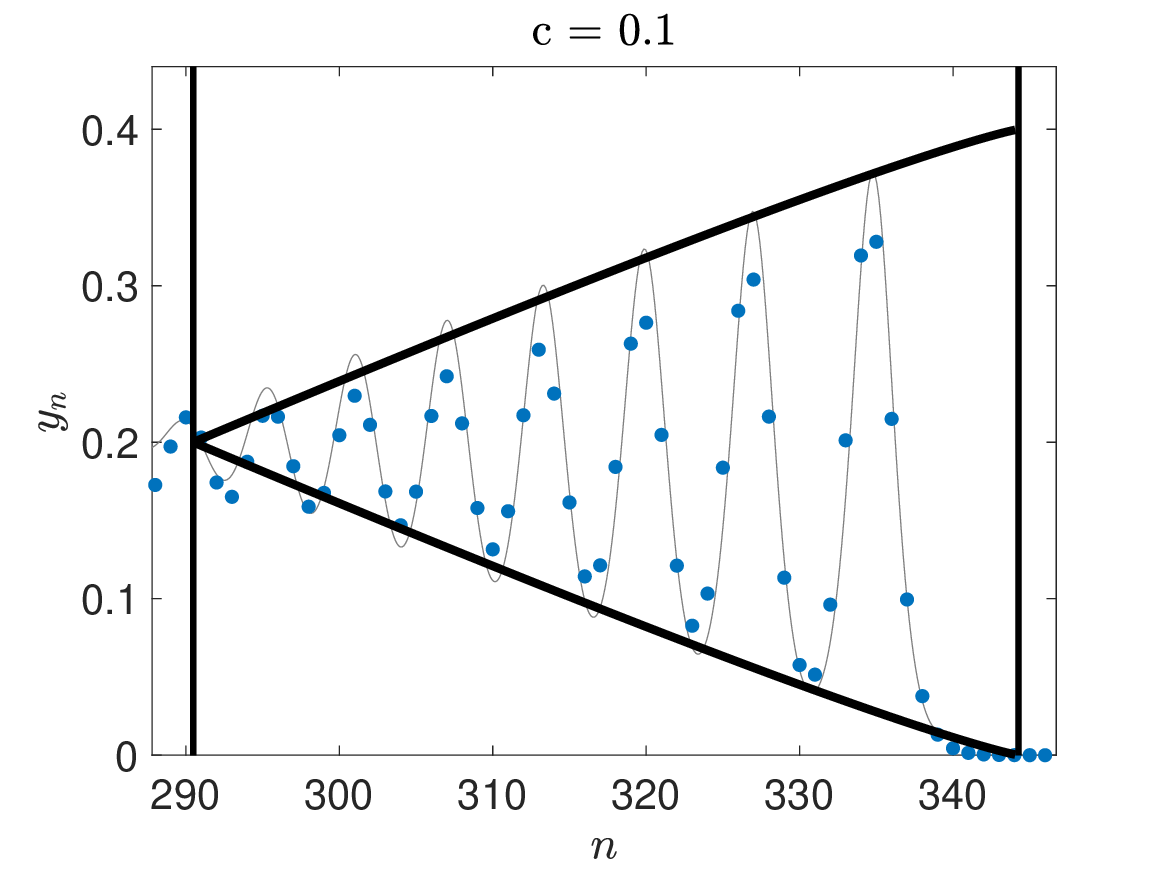} &
  \rlap{\hspace*{5pt}\raisebox{\dimexpr\ht1-.1\baselineskip}{\bf (b)}}
 \includegraphics[height=4.4cm]{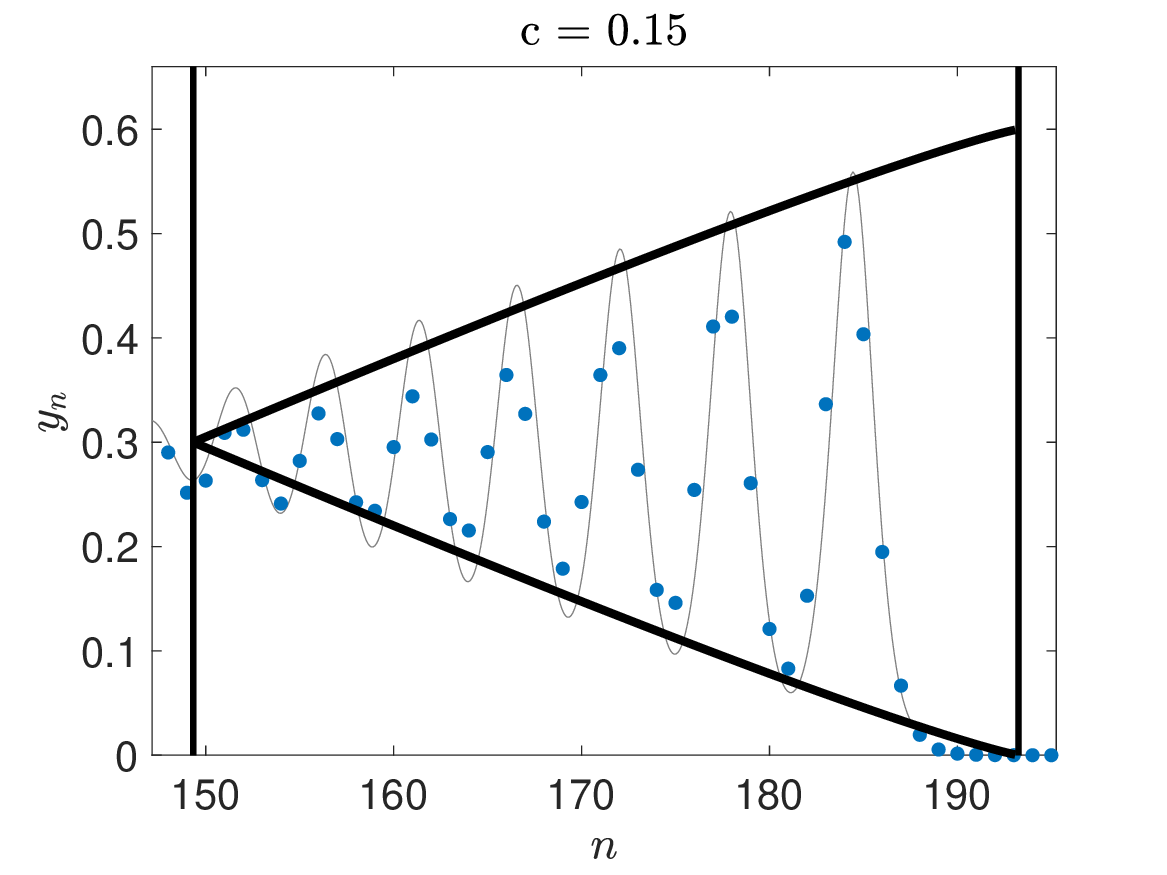} &
   \rlap{\hspace*{5pt}\raisebox{\dimexpr\ht1-.1\baselineskip}{\bf (c)}}
 \includegraphics[height=4.4cm]{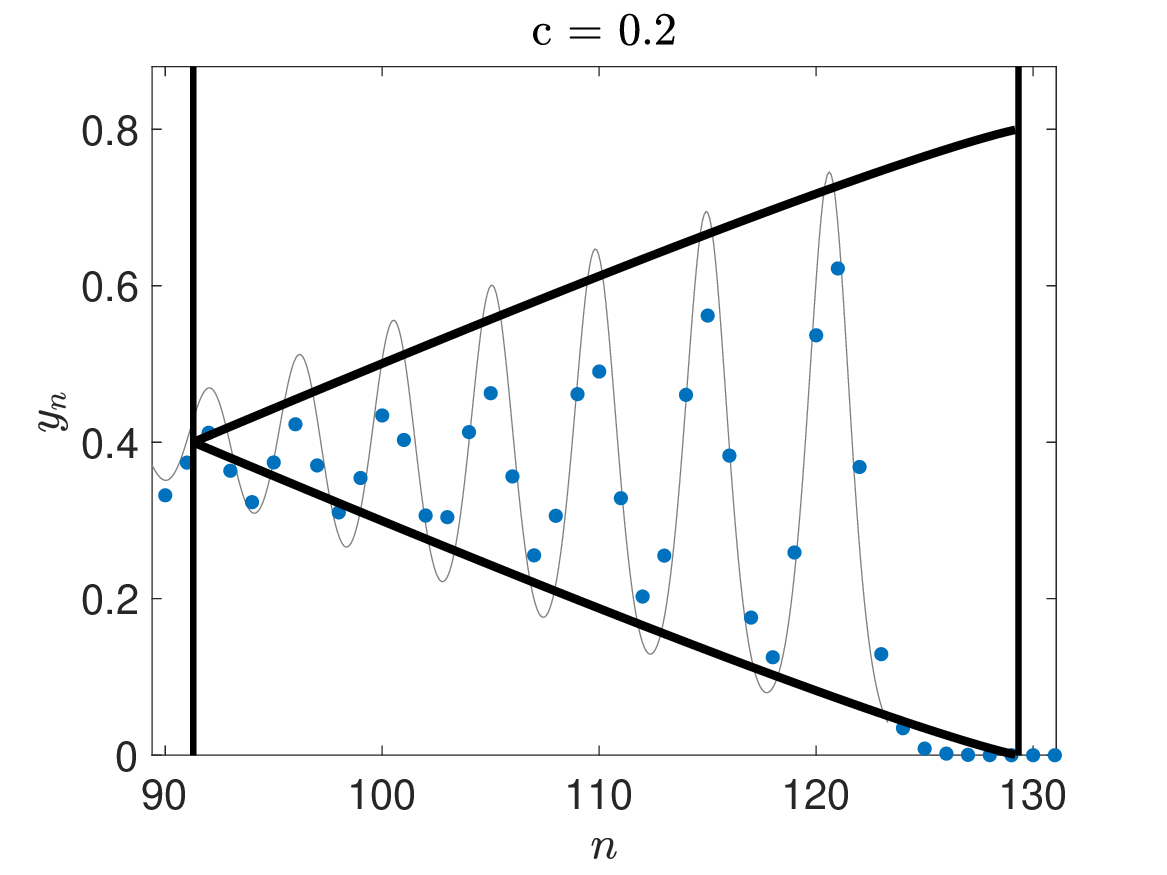} 
  \end{tabular}
  }
  \centerline{
   \begin{tabular}{@{}p{0.33\linewidth}@{}p{0.33\linewidth}@{}p{0.33\linewidth}@{} }
     \rlap{\hspace*{5pt}\raisebox{\dimexpr\ht1-.1\baselineskip}{\bf (d)}}
 \includegraphics[height=4.4cm]{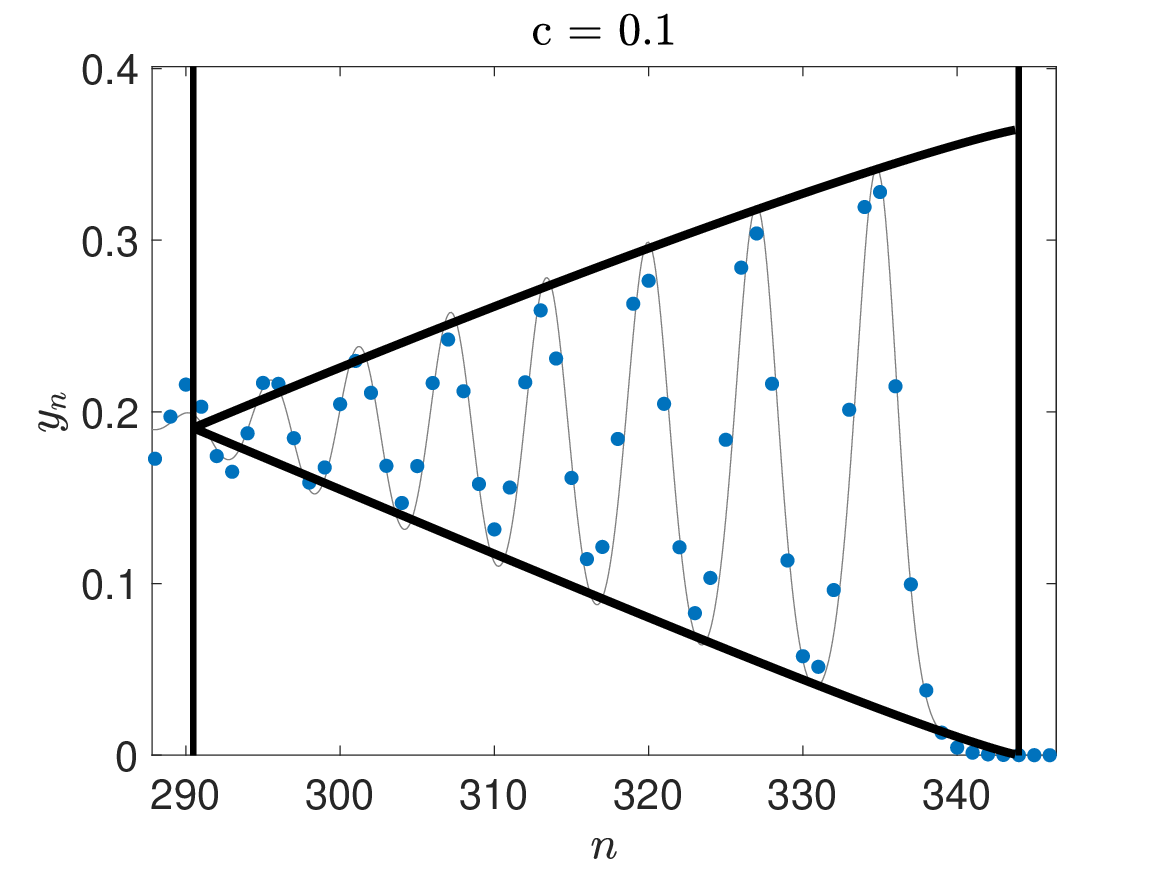} &
  \rlap{\hspace*{5pt}\raisebox{\dimexpr\ht1-.1\baselineskip}{\bf (e)}}
 \includegraphics[height=4.4cm]{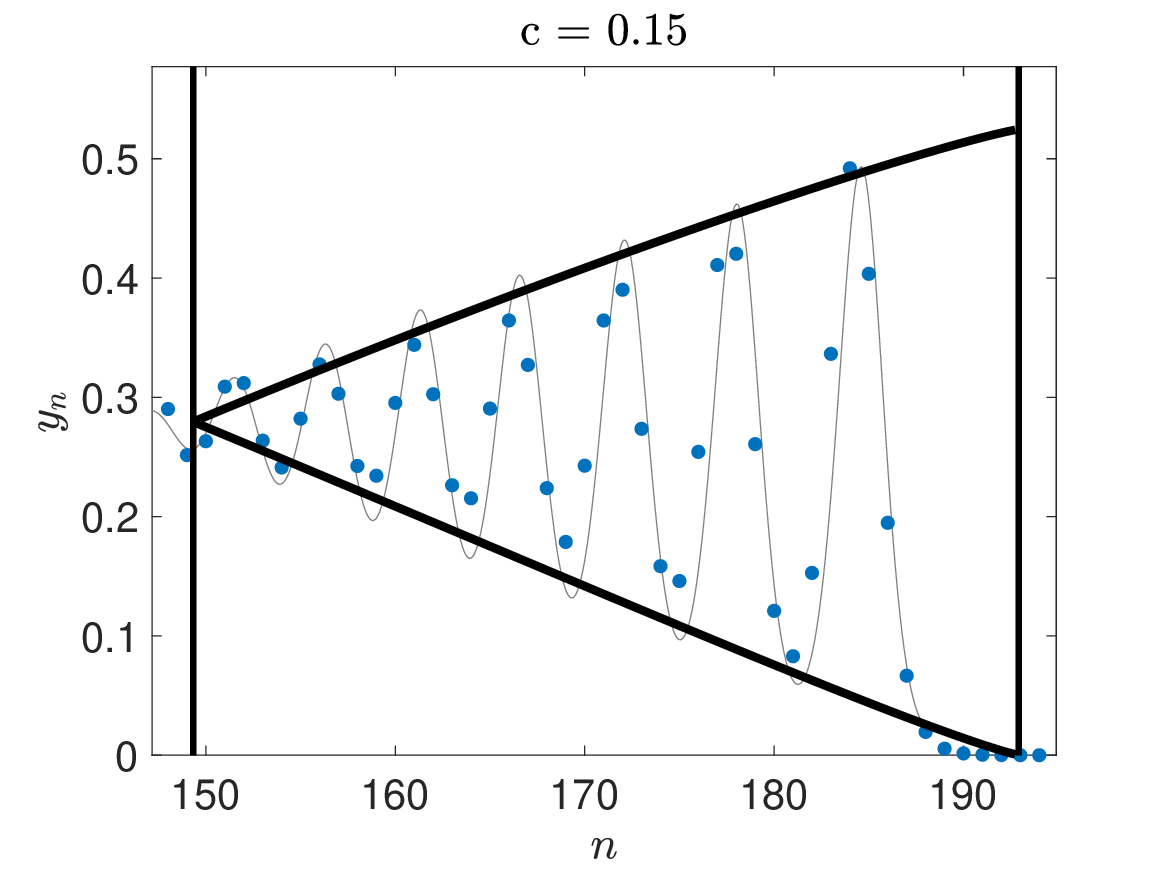} &
   \rlap{\hspace*{5pt}\raisebox{\dimexpr\ht1-.1\baselineskip}{\bf (f)}}
 \includegraphics[height=4.4cm]{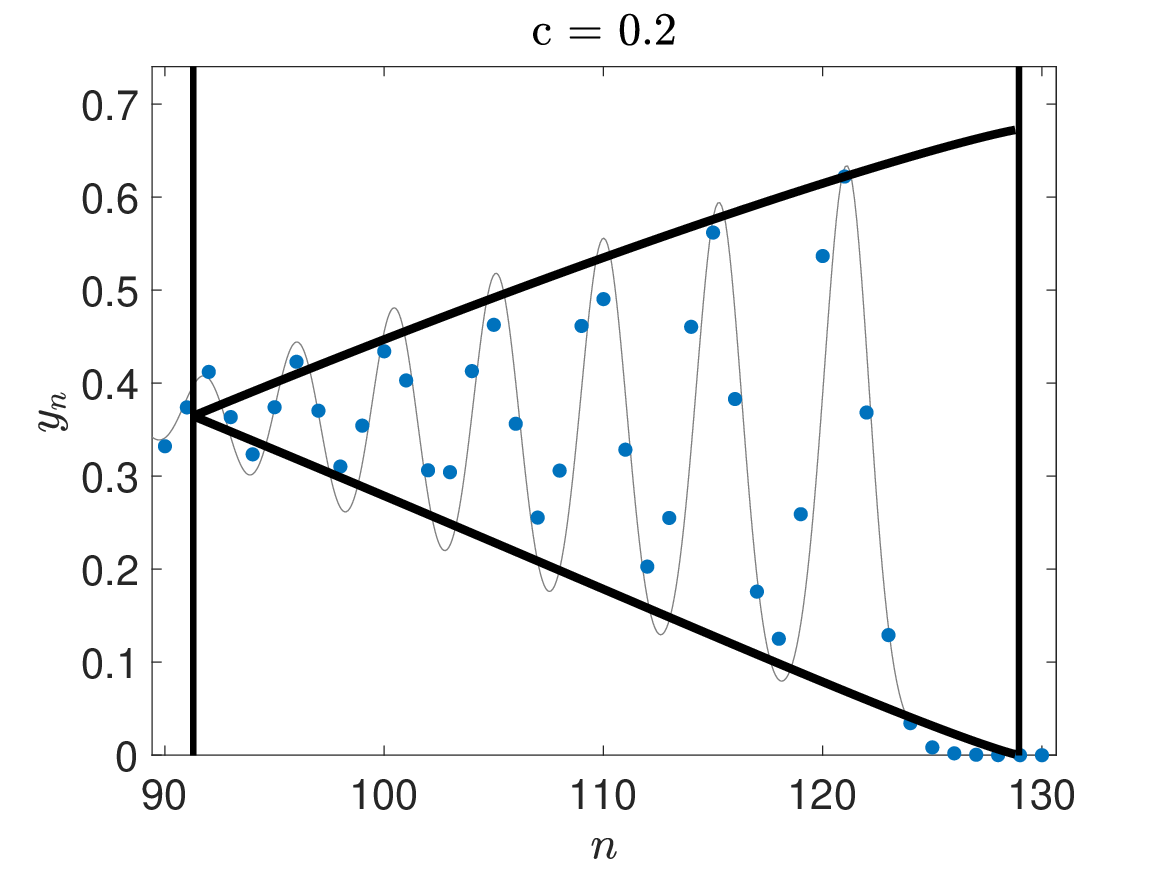} 
  \end{tabular}
  }
    \caption{Granular chain simulations with an initial velocity shock,
    Eq.~\eqref{eq:velocity_shock},  compared to the KdV prediction (top row) and Toda prediction (bottom row). Theoretical predictions are shown as solid lines
    and granular chain simulations are shown as markers.
    In all panels, the variables in the KdV scaling
    are fixed to $T=1200$ and $X\in[-50,30]$. The vertical lines
    are the predictions of the trailing  
    and leading edge. 
    The sloped lines are the prediction of the envelopes.
    \textbf{(a)} The small parameter is $\cd = 0.1$ ($\epsilon \approx 1.55$) and the time in the original granular
    chain scaling is $t \approx 322$. The best-fit phase shift is the linear interpolation
    between $\theta_\ell = -4.5$ and $\theta_r = 0.5$.
    \textbf{(b)} $\cd = 0.15$ ($\epsilon \approx 1.9$), $t \approx 175$,
    $\theta_\ell = -5.5$ and $\theta_r = -0.5$.
     \textbf{(c)}  $\cd = 0.2$ ($\epsilon \approx 2.2$), $t \approx 115$, $\theta_\ell = -6$ and $\theta_r = -1.5$.
\textbf{(d)} $\cd = 0.1$ ($\epsilon \approx 1.55$), $\theta_\ell = -3$ and $\theta_r = 0.2$.
    \textbf{(e)} $\cd = 0.15$ ($\epsilon \approx 1.9$), $t \approx 175$,
    $\theta_\ell = -4$ and $\theta_r = -0.5$.
     \textbf{(f)}  $\cd = 0.2$ ($\epsilon \approx 2.2$), $t \approx 115$, $\theta_\ell = -4.5$ and $\theta_r = -1$.
} 
    \label{fig:velocity_shock}
\end{figure}

\begin{figure}
\centerline{
   \begin{tabular}{@{}p{0.33\linewidth}@{}p{0.33\linewidth}@{}p{0.33\linewidth}@{} }
     \rlap{\hspace*{5pt}\raisebox{\dimexpr\ht1-.1\baselineskip}{\bf (a)}}
 \includegraphics[height=4.4cm]{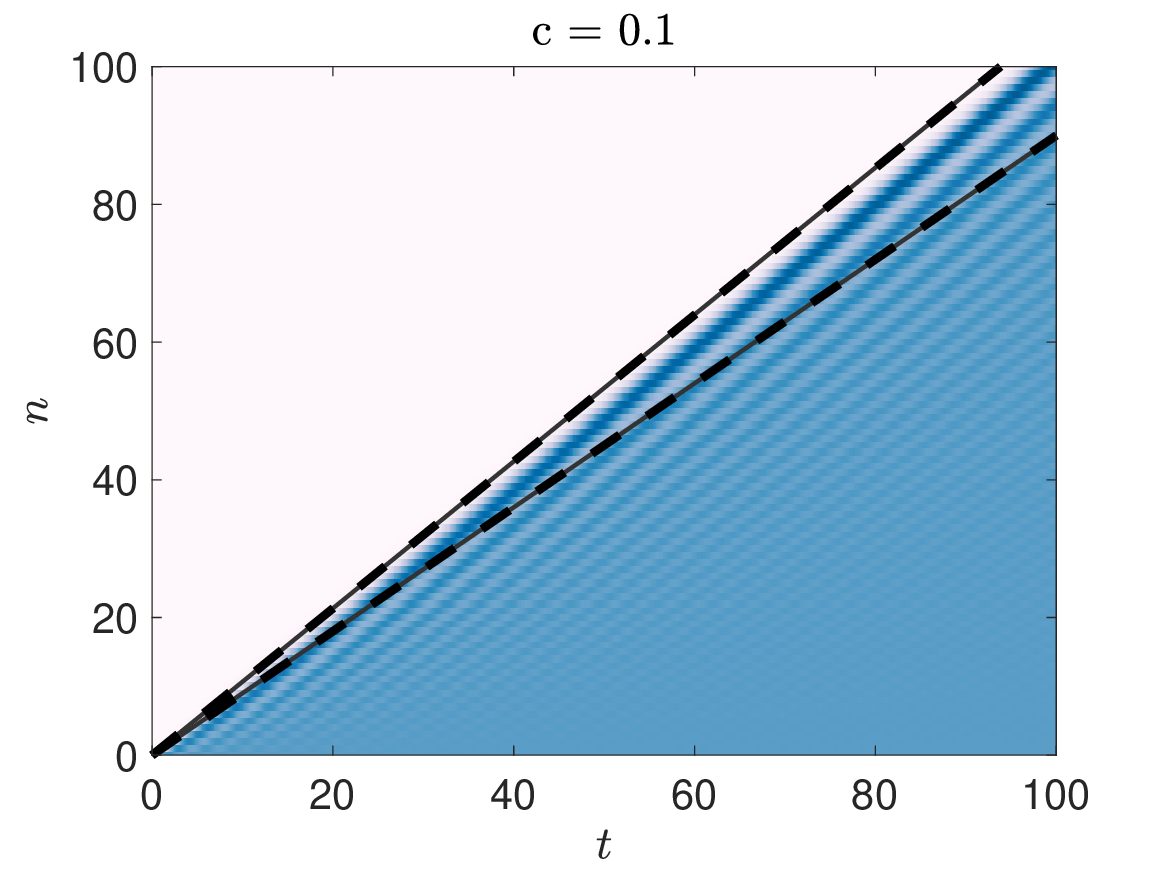} &
  \rlap{\hspace*{5pt}\raisebox{\dimexpr\ht1-.1\baselineskip}{\bf (b)}}
 \includegraphics[height=4.4cm]{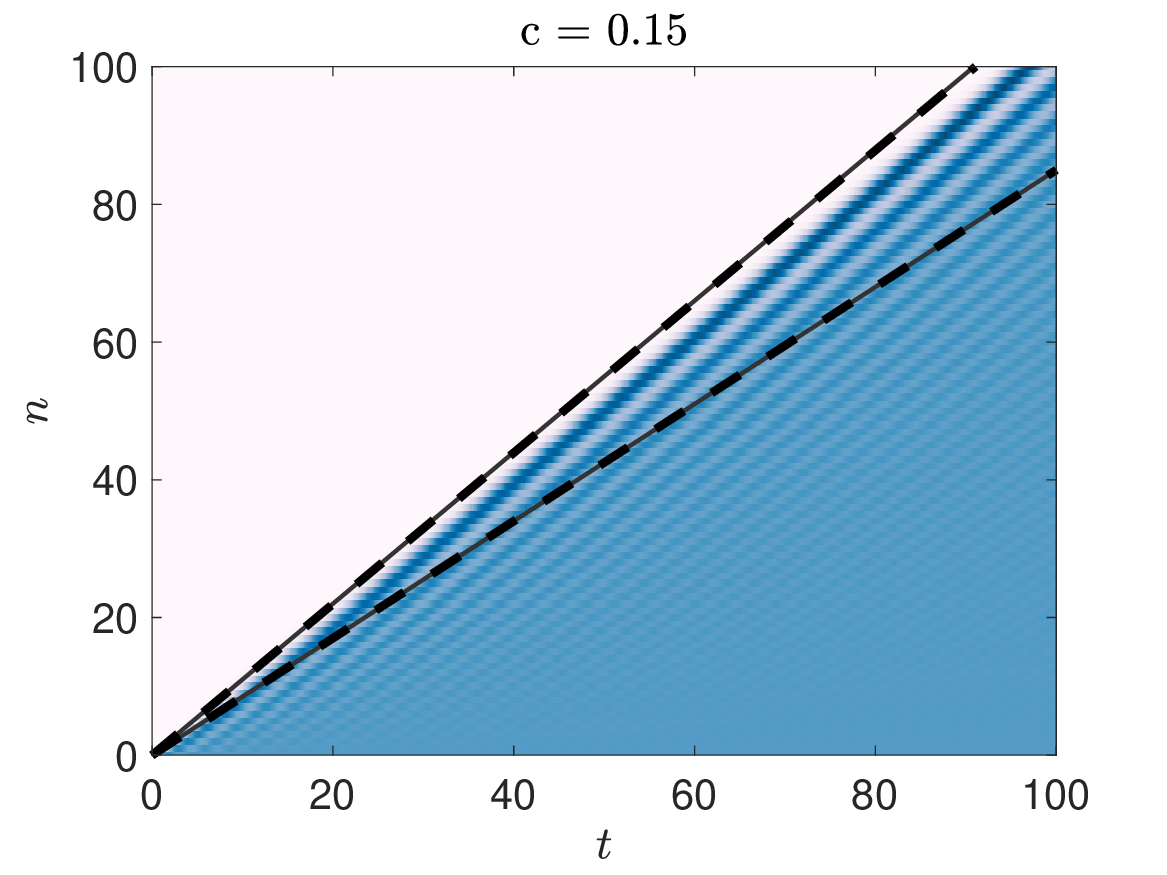} &
   \rlap{\hspace*{5pt}\raisebox{\dimexpr\ht1-.1\baselineskip}{\bf (c)}}
 \includegraphics[height=4.4cm]{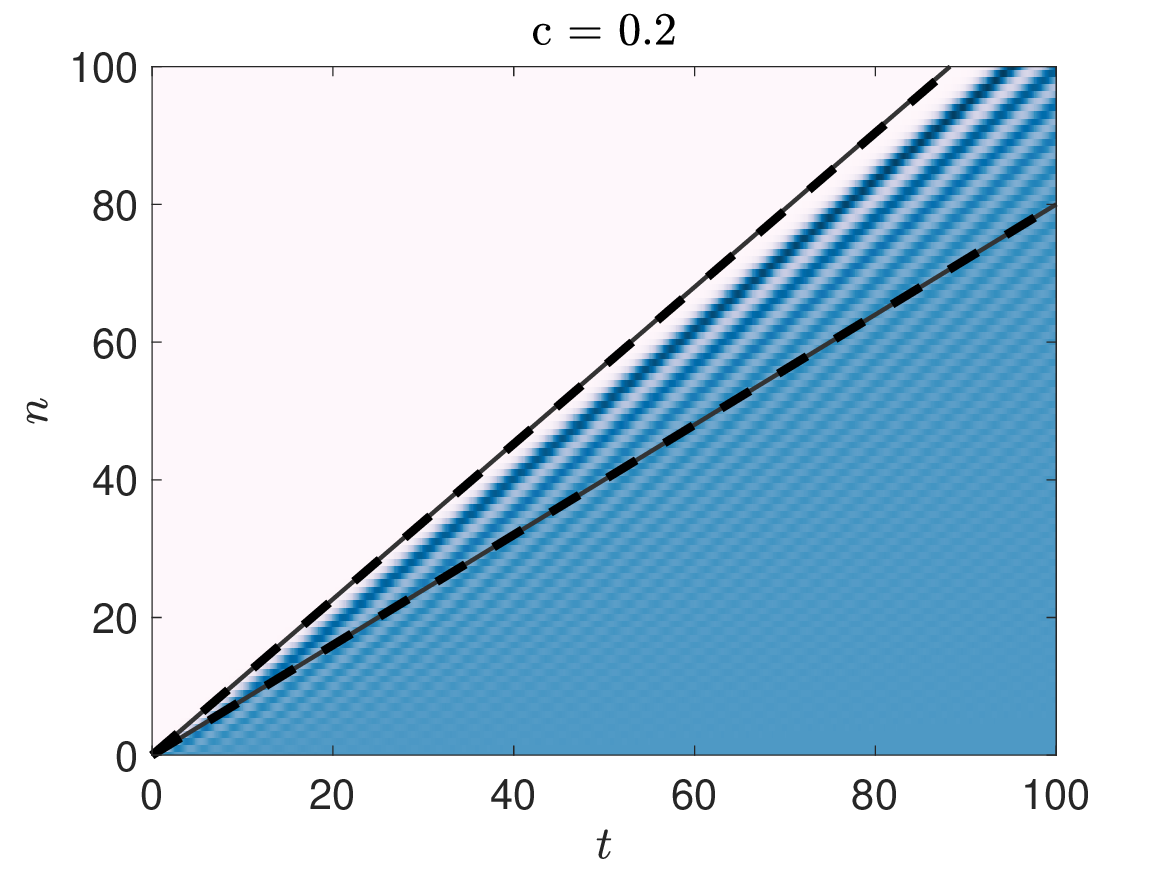} 
  \end{tabular}
  }
    \caption{Intensity plots of the granular chain simulations with an initial velocity shock,
    Eq.~\eqref{eq:velocity_shock}. The solid lines are the estimates of the leading
    and trailing edge from the KdV description, and the dashed lines are from the Toda description.
    \textbf{(a)} $c=0.1$,
    \textbf{(b)} $c=0.15$
     \textbf{(c)} $c=0.2$.
    } 
    \label{fig:velocity_shock_intensity}
\end{figure}

Figure~\ref{fig:kdv_XTfixed} demonstrates that the KdV approximation of the granular
chain DSW is quite good, as long as the parameter $c$ is small. Thinking of the relevance
of the description for an experimental setting, a smaller lattice size would be 
more practically accessible,
and thus the speed $c$ should be larger, such that the DSW can form before
the end of the chain is reached. Thus, we will next explore going beyond
the smallness assumption of $c$ in hopes of getting to lattices of reasonable
size (some experiments have chains on the order of 100 beads~\cite{Boechler2010}
and, indeed,~\cite{khatriprl} reports experiments with up to 188 beads). 
Furthermore,
the initial conditions used for Fig.~\ref{fig:kdv_XTfixed} imply
the entire strain and velocity profile must be specified, which is generally hard
to achieve in granular chain experiments. The more relevant initial condition
for the granular chain, and other similar lattices,  is a collision at
one side of the chain. In \cite{Molinari2009} it was shown that such a collision is
well approximated by a continuous velocity applied to the end of a semi-infinite
chain (this is the so-called piston problem \cite{piston}). By an appropriate change of
variables, the piston-problem initial conditions are equivalent to
an infinite chain with a velocity shock \cite{Venakides99}, namely,
\begin{subequations}
\label{eq:velocity_shock}
\begin{gather}  
      u_n(0) = 0 \\
      \dot{u}_n(0) = - 2 c \, \sign(n)  
\end{gather}
\end{subequations}
where $\sign(0)=0$. Notice that this initial condition is given in terms of the displacement variable $u_n$. Nonetheless,
we continue to report results in terms of the strain $y_n = u_n - u_{n+1}$ for consistency.
With such an initial condition, it is reasonable to suppose,
based on the linear theory \cite{MielkePatz2017}, that the trailing edge (in the strain) will have a mean close to $2c$.This suggests that
the $c$ defined assuming the initial data Eq.~\eqref{step2} is the same
as the $c$ defined in Eq.~\eqref{eq:velocity_shock}. Indeed, this was
the motivation for defining $c=\epsilon^2/24$ previously. This means we can
apply the prediction Eq.~\eqref{eq:granular_kdv_dsw}, even when
the initial condition is given by a velocity shock. The top row
of Fig.~\ref{fig:velocity_shock} shows a comparison of a DSW
formed given a velocity shock (markers) and the KdV prediction
for various values of $c$. 
Once again, the microscopic time $t$ is chosen such 
that macroscopic time $T_f = 1200$ is fixed. This implies
that Fig.~\ref{fig:kdv_XTfixed}(c) and Fig.~\ref{fig:velocity_shock}(a)
show the same KdV approximation, but the former granular chain data
comes from a smooth shock in the strain, while the latter has
a (displacement) velocity shock. By comparing those two figure panels, one can
see that the overall structure of the two DSWs is qualitatively similar.
Figure~\ref{fig:velocity_shock}(b,c) show examples for larger values of $c$,
and the discrepancies are becoming more evident. Nonetheless
the qualitative agreement is still reasonable. Notice
that for $c=0.2$, which corresponds to Figure~\ref{fig:velocity_shock}(c),
a chain extending to $n=130$ will capture the DSW, which is within
the realm of current experimental capabilities, as per our
discussion of~\cite{khatriprl} above.  The top
row of Fig.~\ref{fig:velocity_shock_intensity} shows the same simulation
data, but now with windowing such that the microscopic variables ($n,t$)
are fixed. In particular, intensity plots of the strain are shown.
The prediction of the leading and trailing speed from the KdV equation
are shown as solid black lines.

\begin{figure}
\centerline{
   \begin{tabular}{@{}p{0.5\linewidth}@{}p{0.5\linewidth}@{} }
     \rlap{\hspace*{5pt}\raisebox{\dimexpr\ht1-.1\baselineskip}{\bf (a)}}
 \includegraphics[height=6.5cm]{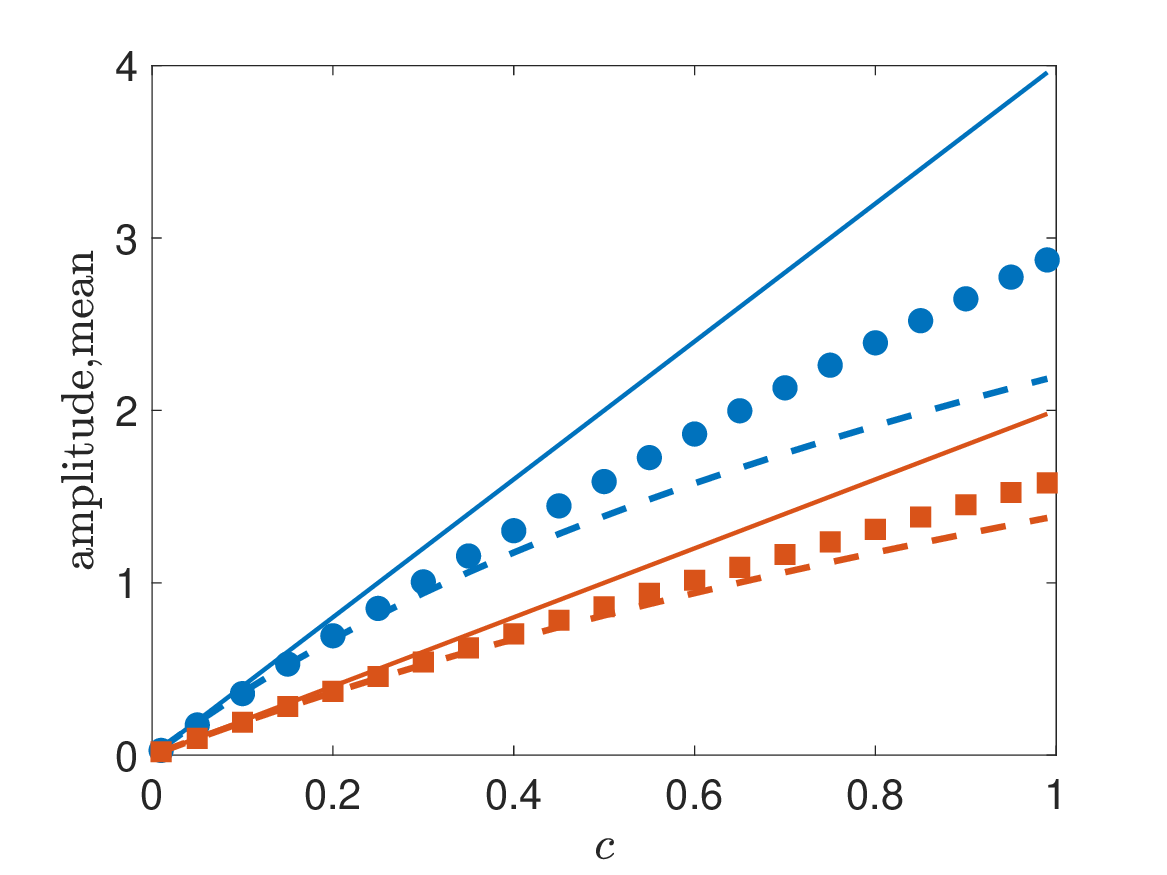} &
   \rlap{\hspace*{5pt}\raisebox{\dimexpr\ht1-.1\baselineskip}{\bf (b)}}
 \includegraphics[height=6.5cm]{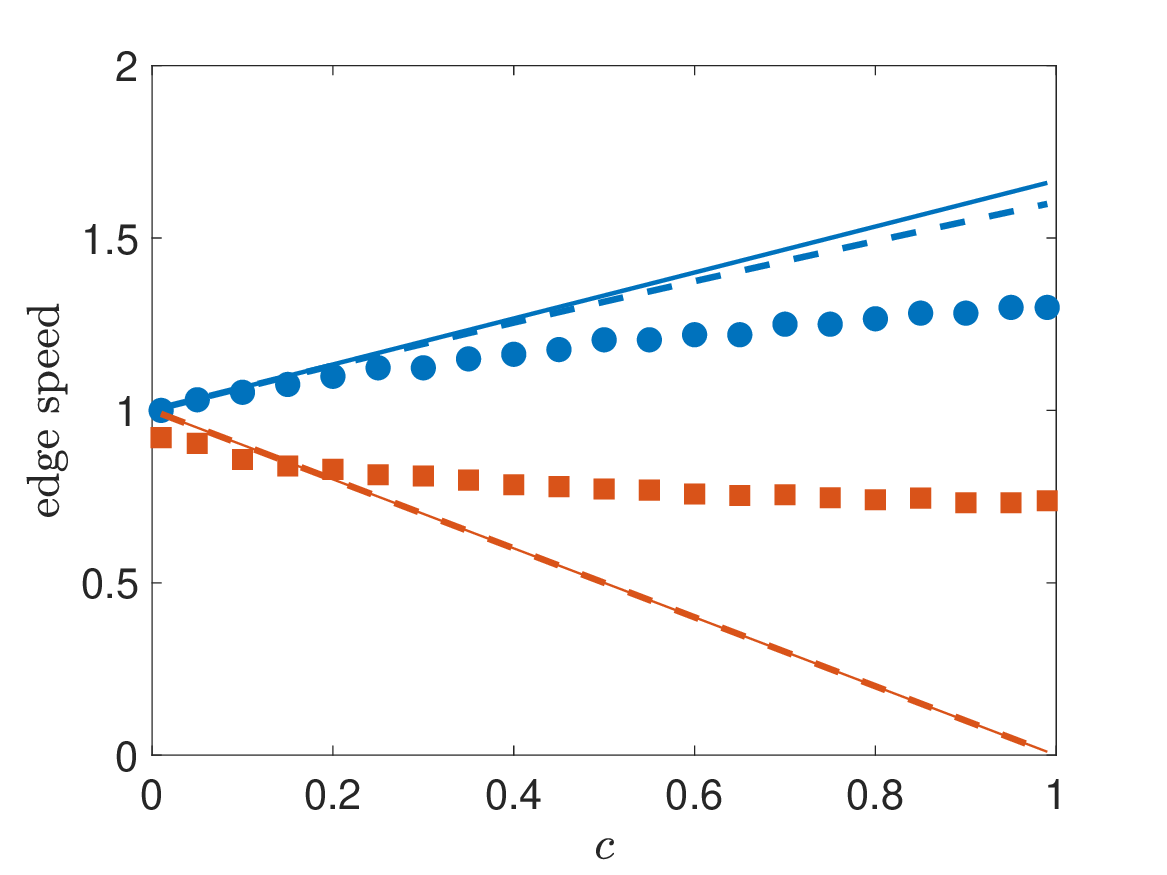} 
  \end{tabular}
  } 
\caption{\textbf{(a)} Numerical estimates  of the leading edge amplitude (blue dots)
and trailing edge mean (red squares) of the granular DSW versus the initial velocity parameter $c$.
The corresponding KdV approximation (solid lines) and Toda approximation (dashed lines)  
are also shown.
\textbf{(b)} Numerical estimates  of the leading edge speed (blue dots)
and trailing edge speed (red squares) of the granular DSW versus the initial velocity parameter $c$.
The corresponding KdV approximation (solid lines) and Toda approximation (dashed lines)  
are also shown.
}
\label{fig:toda_amp_speed}
\end{figure}

We now investigate how well the estimates of the trailing and leading
edge characteristics in Eq.~\eqref{eq:kdv_speeds_param} compare to the granular DSWs for a
larger range of the parameter $c$. Figure~\ref{fig:toda_amp_speed}(a)
shows a numerical estimate of the leading edge amplitude
(blue dots). The amplitude
is simply the maximum strain observed at the final time
of the simulation, which we fixed to $t_f = 900$ for all simulations
shown in the figure.  The KdV
approximation of the leading edge amplitude, $a_+^{\rm KdV}$,
is shown as the blue solid line. As expected, the agreement is good
for small $c$, but gradually gets worse as $c$ increases.
The trailing edge mean of the simulated DSW is shown
as the red squares in Figure\ref{fig:toda_amp_speed}(a). It
is estimated as the mean of the first node, namely $\frac{1}{T}\int_{I_T} y_1(t) dt$
where $T$ is the peak to peak time of the final oscillation in the simulation
and $I_T$ is the corresponding interval of time. The 
KdV approximation, $\bar{y}_-^{\rm KdV}$, is shown as the solid red line.
The leading edge (blue dots) and trailing edge (red squares) speeds 
are shown in 
Fig.~\ref{fig:toda_amp_speed}(b).
The leading speed is estimated by computing the difference
in the times the maximum is obtained between two consecutive sites
and simply inverting that time difference to obtain the speed estimate.
We used the sites $n=700$ and $n=701$. To estimate the trailing edge speed,
we select a small amplitude threshold and define the trailing edge
to be the first node that achieves a strain higher than or equal
to the threshold after having fixed the time to $t=t_f$. The estimate
for the speed is then simply this critical lattice site divided by $t_f$. One must define
such a threshold since there will always be the presence of small
amplitude linear waves at the trailing edge of the DSW.  The
threshold we used was the mean of the tail (just described)
plus 2.5\% of the maximum strain of the DSW. This particular
choice of threshold yields good agreement when numerically
comparing the trailing speed of the Toda lattice DSW
to the analytical prediction (detailed in the next section).
The KdV prediction of the edge speeds are shown as the solid
lines for the trailing edge, $s_-^{\rm KdV}$ (red),
and leading edge, $s_+^{\rm KdV}$ (blue).
From Fig.~\ref{fig:toda_amp_speed} we see that the KdV prediction
overestimates the leading edge amplitude and trailing edge mean,
and overestimates (by quite a large margin for large $c$)
the DSW core length, given by $(s_+^{\rm KdV}-s_-^{\rm KdV})t$.

\section{Toda Description of the Granular Chain} \label{sec:toda}

We now turn our attention to a different analytical approximation of the granular chain DSW.
The latter will be based on the Toda lattice, which is one of the few nonlinear lattices that is integrable (another
example is the Kac-van Moerbeke lattice, which is closely related to Toda
via a Miura type transformation \cite{Moser}).
The Toda lattice has a rich history~\cite{Toda}. The equation in its
canonical form is
\vspace*{-0.4ex}
\begin{equation}
   \ddot u_n = e^{u_{n-1}-u_n} - e^{u_n-u_{n+1}}\,.
\label{e:Toda_d2ydt2}  
\end{equation}

If one Taylor expands the relevant ODEs, the following FPUT lattice is obtained,
\begin{equation} \label{eq:fput_toda}
	\ddot{u}_n = (u_{n-1} - 2 u_n + u_{n+1}) 
                 -\frac{1}{2}\left[ (u_{n+1} - u_{n-1})^2 - (u_{n} - u_{n-1})^2   \right] + 
                       \frac{1}{6}\left[ (u_{n+1} - u_{n-1})^3 - (u_{n} - u_{n-1})^3   \right] 
\end{equation}
Through the rescaling $u_n(t) \rightarrow K_2/K_3 u_n( t  \sqrt{K_2})$, we can match the linear and quadratic terms of the Taylor-expanded granular chain (see Eq.~\eqref{eq:fput_granular}) and 
the Taylor-expanded Toda chain (see Eq.~\eqref{eq:fput_toda}) for an arbitrary choice of the granular chain parameters $d_0,\gamma,M$ (recall $p=3/2$).
Since we can only rescale time and amplitude, we cannot match the cubic coefficients. 
Notice that a similar matching was leveraged in the work of~\cite{Shen}
in order to study traveling waves in granular chains: indeed, in that connection
the Toda lattice could afford the possibility of bi-directional waves and their
collisions,
a feature that is absent from the (unidirectional) KdV approximation. 
Recall that for convenience we selected
$\gamma = 2^{3/2}/3$, $M=1$, $d_0 = 1/2$
for the numerical computations, which
led to Taylor coefficients
 $K_2 = 1$ and $K_3 = -1/2$, which match the first two Toda Taylor coefficients.
This was the reason for the choice of parameters used in the previous section,
so that we may better compare the KdV and Toda predictions.
Recall that the sound speed is $\sound = 1$.

There is a four parameter family of traveling periodic waves of the Toda lattice,
parameterized by $E_1,E_2,E_3,E_4$.  In terms of the strain, the formula is,
\begin{equation}
y_n(t) = \log\left( 2 \hat R_n(t) + \frac{1}{2}(E_1^2 + E_2^2 + E_3^2 + E_4^2) - \mu_n^2(t) - b_n^2(t)  \right)
\label{e:aelliptic}
\end{equation}
where
\begin{equation} \label{e:belliptic}
    b_n(t) = E_1 + E_2 + E_3 + E_4 - 2\mu_n(t)
\end{equation}
and
\bse
\begin{gather}
\mu_n(t) = E_2 \frac{\displaystyle 1 - (E_1/E_2)\,B\,\sn^2(Z_n(t),m)}{1 - B\,\sn^2(Z_n(t),m)}\,,
\\
Z_n(t) = 2n F(\Delta,m) + \omega t\ + Z_0, \label{e:Z}
\\
m = \frac{(E_3-E_2)(E_4-E_1)}{(E_4-E_2)(E_3-E_1)}\,,
\label{e:mdef}
\\
\hat R_n(t) = -\sigma_n(t) \,\sqrt{P(\mu_n(t))}\,,\qquad
P(z) = (z-E_1)(z-E_2)(z-E_3)(z-E_4)\,,
\\
\omega = \sqrt{(E_4-E_2)(E_3-E_1)}\,,\qquad
\Delta = \sqrt{\frac{E_4-E_2}{E_4-E_1}}\,,\qquad
B = \frac{E_3-E_2}{E_3-E_1}\,.
\end{gather}
\ese
In the above equations,
$\sn(z,m)$ is the Jacobi elliptic sine, 
$F(z,m)$ is the inverse of $\sn(z,m)$, 
$Z_0$ is an arbitrary translation parameter (phase), $\mu_n(t)$ is the Dirichlet eigenvalue of the scattering problem for the Toda lattice
and $\sigma_n(t) = \pm1$ is the sign associated with $\mu_n(t)$, and determines whether $\mu_n(t)$ is increasing or decreasing as a function of $n$ and $t$. For numerical computations,
we used $\sigma_n(t) = -\sign( \mod(Z_n(t)/K_m) - 1/2   )$.
Notice how the Toda traveling wave solution is more complicated
than its KdV counterpart in Eq.~\eqref{per}. So while we may anticipate a better approximation
(since no long-wavelength assumption is made), the cost is a formula that will be more cumbersome.

\subsection{DSWs of the Toda lattice}
The Toda shock problem \cite{PRB1981v24p2595,VDO}
is the IVP for Eq.~\eqref{e:Toda_d2ydt2} with an initial velocity
shock, see Eq.~\eqref{eq:velocity_shock}.
Both the shock problem and the rarefaction problem were studied in \cite{blochkodama} 
within the framework of Whitham modulation theory. Like in the KdV case,
one can derive a system of modulation equations (4 in the Toda case) that can be written in diagonalized form, and solved in the self-similar frame $E_j=E_j(n/t)$ assuming an initial velocity shock. Three of the parameters
are constant in the self-similar frame, and one depends on the parameter
$m$ \cite{blochkodama,Gino2023}. In particular,
\begin{equation} \label{eq:Es}
    E_1  = -(1+c), \quad
E_2  = -(c-1), \quad 
E_3(m) =   (c+1) \frac{1-c(1-m)}{1+c(1-m)}, \quad
E_4 = (c+1). 
\end{equation}
Assuming $0<c<1$, the core of the DSW
is described by Eq.~\eqref{e:aelliptic} with parameters given via
Eq.~\eqref{eq:Es}
where $n,t$ are parameterized by $m$ through the expression
\begin{equation} \label{eq:Toda_vel}
\frac{c(c+1)}{1+c(1-m)}
  \frac{(1+c(1-m)) E(m) - (c+1)(1-m) K(m)}{(c+1) K(m) - (1+c(1-m))\Pi \left(\left.\frac{cm}{c+1}\right|m\right)} =: s^{\rm Toda}(m)  = \frac{n}{t}  
\end{equation}
where $K(m)$, $E(m)$ and $\Pi(n|m)$ are complete elliptic integrals of the first, second, and third kind respectively, and $s^{\rm Toda}(m)$
is the third charecteristic velocity of the Whitham equations for the
Toda system. As before, $m\rightarrow 0$ corresponds to the trailing, harmonic wave edge (recall $0<c<1$) and 
$m\rightarrow 1$ corresponding to the leading, solitary wave edge.
The trailing edge speed ($s_-^{\rm Toda}$) and leading edge speed ($s_+^{\rm Toda}$)  are obtained as limiting values in Eq.~\eqref{eq:Toda_vel}, in particular
\begin{equation} \label{eq:Toda_edgespeeds}
 \lim_{m\rightarrow 0} s^{\rm Toda}(m) =s_-^{\rm Toda} = 1-c, \qquad \lim_{m\rightarrow 1} s^{\rm Toda}(m) = s_+^{\rm Toda} =  \frac{\sqrt{c(c+1)}}{\log(\sqrt{c}+\sqrt{c+1})}
\end{equation}
Once again, we have that the leading edge speed is supersonic whereas the trailing edge speed is subsonic. 
We can also compute the trailing edge mean and leading edge amplitude from Eq.~\eqref{e:aelliptic},
\begin{equation} \label{eq:Toda_meanamp}
\bar{y}_-^{\rm Toda} = 2\log(1+c), \qquad a_+^{\rm Toda} = 2 \log(1 + 2c)
\end{equation}
Notice that trailing edge speed predictions from Toda and KdV are identical, namely $s_-^{\rm Toda} = s_-^{\rm KdV}$.
Indeed, the remaining three edge characteristics are also related. By
Taylor expanding the remaining formulas in Eqs.~\eqref{eq:Toda_edgespeeds} and \eqref{eq:Toda_meanamp} about $c=0$ shows that the leading order behavior is identical to the KdV approximation, shown in Eq.~\eqref{eq:kdv_speeds_param}. Namely, for small $c$
we have that
\begin{equation} \label{eq:compare_parms}
s_+^{\rm Toda} \approx 1 + \frac{2}{3} c = s_+^{\rm KdV}, \qquad \bar{y}_-^{\rm Toda} \approx 2c = \bar{y}_-^{\rm KdV},\qquad
a_+^{\rm Toda} \approx 4c = a_+^{\rm KdV} 
\end{equation}

\subsection{Comparison of Toda and granular DSWs}

We now compare the Toda predictions of the granular DSW via Eq.~\eqref{e:aelliptic} (with parameters defined in Eq.~\eqref{eq:Es})
with direct simulations of the granular chain with the velocity shock initial data, defined in Eq.~\eqref{eq:velocity_shock}.
The bottom row
of Fig.~\ref{fig:velocity_shock} shows the simulation (markers) and the Toda prediction (lines)
for various values of $c$. In order to make concrete comparisons with the KdV predictions
(shown in the top panels of this figure),  the time is chosen such 
that $T_f = 1200$ is fixed. This implies that the top panels can be directly
compared with the panel beneath it. Note that the granular chain 
simulation data is identical in both
cases (the markers) and only the analytical predictions differ.
Qualitatively, the KdV predictions are quite similar to the Toda ones.
From a quantitative perspective, one can clearly see that Toda performs
better as $c$ gets larger
(compare Fig.~\ref{fig:velocity_shock}(b,c) to panels (e,f)).
Figure~\ref{fig:velocity_shock_intensity} shows the same simulation
data, but now with windowing such that the microscopic variables ($n,t$)
are fixed. In particular, intensity plots of the strain are shown.
The prediction of the leading and trailing speed from the Toda equation
are shown as dashed black lines. Trailing edge speeds from each
prediction overlap perfectly, since the approximations are identical,
and one can see that the leading edge speed is slightly smaller in
the Toda case.

The estimates of the trailing and leading
edge characteristics in Eq.~\eqref{eq:Toda_edgespeeds} and Eq.~\eqref{eq:Toda_meanamp}
are shown as the dashed lines in Fig.~\ref{fig:toda_amp_speed}.
In particular, in panel (a), the blue dashed line
is the leading edge amplitude $a_+^{\rm Toda}$ and the red dashed
line is trailing edge mean  $\bar{y}_-^{\rm Toda}$. Note
that in each case, the Toda prediction underestimates the relevant
quantities, whereas the KdV prediction overestimates them. As 
expected from Eq.~\eqref{eq:compare_parms}, both the KdV
and Toda predictions become closer as $c\rightarrow 0$, and that
both get closer to the actual granular DSW dynamics.
The blue dashed line in panel (b) is the Toda
prediction of the leading edge speed $s_+^{\rm Toda}$.
Here, the Toda prediction is once again smaller than
the KdV prediction, however both in this case are overestimates
of the actual granular DSW leading edge speed. Both estimates
become more accurate as $c$ become smaller. The trailing edge
speed predictions are identical in the KdV and Toda case,
which underestimate the actual trailing edge speed.
The asymptotic prediction for $c\rightarrow 0$ is 
not captured, but this is likely due to the error involved
in numerically estimating the trailing edge speed due
to the presence of small amplitude linear waves
present in the simulations. These tails only vanish
for $t\rightarrow \infty$.

Finally, we remark that we have only considered values of $0<c<1$,
which correspond to the Genus-2 region of the Whitham equations
for the Toda lattice. In this subcrtical case, the parameter
$m$ has a minimum value of $m=0$ and the tail of the solution
thus approaches a zero amplitude constant. In the supercritical
case of $c>1$, the corresponding Whitham equations are in the Genus-1
region, and the minimum value of $m$ is $m=1-\frac{1}{c^2}$. In this case
the inner part of the DSW (outside of the core of the DSW) is a binary
oscillation \cite{blochkodama}. This is because the wavelength has reached,
at this critical value of $m$, an integer value such 
that the oscillations, in the lattice, are binary.
In the subcritical case, the wavelength does not reach unity
before the end of the core is met (when traversing the DSW from the
leading edge to trailing edge).
While the KdV and Toda predictions are clearly outside
their range of validity for $c>1$, it is worth noting that
we did not observe any bifurcation of the tail behavior transitioning
from a zero amplitude constant to a binary oscillation (we tested
up to $c=100$). In other words, the wavelengths of oscillation near the tail were
always less than unity for the values of $c$ we tested.
This observation is especially interesting given that
the Taylor expanded approximation of Toda (or the granular chain), i.e.,
the FPUT chain,
does indeed undergo such a bifurcation \cite{Venakides99}.
The lack thereof
seems to be a feature of the nonlinearity of the granular chain.
This observation is a key difference between the granular chain DSWs
and their weakly nonlinear approximations and it underscores the importance
of investigations in the strongly nonlinear regime. 
Such studies will be reported on in future publications. 

\section{Conclusions and future challenges}
\label{s:conclusions}

In the present work we have revisited the shock wave problem of
a system of wide relevance to material science applications, namely
the granular chain. Such nonlinear dynamical lattices in both their
monomer, and even in their heterogeneous (e.g., dimer) installments
have been explored not only theoretically but also in physical 
experiments over the last two decades. Recently, additional related
settings such as tunable magnetic
lattices, have also become experimentally accessible~\cite{talcohen}.
Importantly, in all of these settings the advances of experimental
monitoring technology enables the real-time visualization of
structures including traveling~\cite{granularBook} and dispersive
shock waves~\cite{talcohen}. While traveling waves have been 
the quintessential workhorse of such lattices and of their applications,
recent analytical, numerical and experimental developments have
motivated the systematic consideration of DSW structures.

Herein, we have leveraged the approximation of the granular chain
by a Fermi-Pasta-Ulam-Tsingou lattice, and at a second layer of
approximation of the latter by an integrable continuum (the KdV)
or discrete (the Toda lattice) system. The fundamental advantage
of these integrable settings here is not their analytical
solvability via the inverse scattering transform, although certainly
that might be desirable. Rather, it is instead their accessibility,
through seminal works such as~\cite{GP73} (for the KdV) and~\cite{blochkodama}
(for the Toda case) of the analysis through Whitham modulation theory
of the DSW patterns. The explicit form of the periodic solutions
of these models lends itself to the slow modulation of their 
(amplitude, width, wavenumber, frequency, etc.) characteristics
which Whitham modulation theory is eminently suited to describe.
Naturally, there are still empirical selections (such as the phase
one made herein), but nevertheless, we have found that in the appropriate
small amplitude regime, these approximations provide an excellent analytical
handle on the form of the DSWs of the granular chain. Indeed, at some level,
given the layers of approximation (granular to FPUT, then FPUT to KdV or Toda),
this may seem somewhat surprising, however this turns out to be a useful
description of precompressed granular chains, as it did earlier also for
the respective traveling waves~\cite{Shen}. Naturally, as the amplitude
of the wave grows, the quality of the approximation decreases. Among the
two approximations the elegant KdV one is simpler, but typically overestimates
more the quantities of the granular chain, while the more cumbersome Toda
one avoids the long wavelength (extra layer of) approximation and 
performs better in larger amplitude settings.

Naturally, such studies pave the way for further/deeper exploration 
of DSWs in discrete systems, an intriguing and quite active 
area of investigation~\cite{wilma,talcohen,CHONG2022133533,Gino2023}.
For instance, one can envision developing quasi-continuum
(possibly regularized) approximations of the original granular chain and
developing a Whitham modulation theory for the resulting continuum model.
One could also imagine developing a modulation theory at the level
of the original discrete problem (alone lines of thought similar
to those, e.g., of~\cite{CHONG2022133533}). Finally, while most
of these studies have been focused on one-dimensional lattices,
both optical applications~\cite{fleischer2} and magnetic ones~\cite{Chong_2021}
suggest the relevance of corresponding explorations in higher (e.g., two-) dimensional
discrete systems.

\section*{Acknowledgments}
This work is dedicated to the memory of Professor Noel F. Smyth,
with gratitude also to the Editors of the associated 
special issue of Wave Motion
for their efforts and kind invitation.
The authors would like to thank the Isaac Newton Institute for Mathematical Sciences for support and hospitality during the programme Dispersive Hydrodynamics when work on this paper was undertaken (EPSRC Grant Number EP/R014604/1).
This material is also based upon work supported by the US National Science Foundation under Grants DMS-2009487 (G.B.), DMS-2107945 (C.C. and A.G.), DMS-2204702 (P.G.K.) and PHY-2110030 (P.G.K).

\bibliographystyle{unsrt}
\bibliography{Chong_Toda}

\end{document}